# The Vision and the Perspective of Digital Tourism

Olga Kononova[1,3[0000-0001-6293-7243]], Dmitry Prokudin[1,2,3[0000-0002-9464-8371]] and Julia Ryabysko[1[0000-0003-3122-2805]]

[1] Saint-Petersburg National Research University of Information Technology, Mechanics and Optics, Kronverksky pr., 49, 197101, Saint Petersburg, Russia
[2] Saint-Petersburg State University, Universitetskaya Emb., 7/9, 199034, Saint Petersburg, Russia
[3] Center digital society research, Russia
kononolg@yandex.ru, hogben.young@gmail.com, u.rbs@yandex.ru

**Abstract.** The dynamics of the modern information society changes the usual areas of human activity, generates various innovations based on the widespread use of Information and Communication Technologies (ICTs). Virtually, every activity today is technology related. In these conditions, scientific activity is also changing. Digitalization processes act as integrative to various scientific directions, which form the base for interdisciplinary scientific research. The study of their formation is an important scientific task aimed at predicting the development of both science and society as a whole. In this study, based on the integrated use of ICTs, we consider methods of the terminology base analysis in various interdisciplinary research directions on the instance of tourism in the digital age. The development of scientific interest in the area of digital tourism in Russian and global scientific discourses is also compared. The purpose of this paper is to proof the relevance of scientific study in the field of tourism digitalization, to identify the generic directions and trends of digital tourism, and to specify technologies for the implementation of digital tourism using the case study of St. Petersburg.

## 1. Introduction

The current level of digitalization of society suggests new forms of interaction between producers and consumers, including in the area of tourism services [5]. Manufacturers of tourism services are forced to introduce modern digital technologies, thereby forming a new direction – digital tourism or tourism digitalization within more general direction – digital economy. Although, there is no any unified definition of the term 'digital tourism' yet, scientific and media discourse offers some approaches and a vision of this phenomenon through the digitalization prism. As a result, the primary tasks of modern tourism are formed and reflected in "The strategy of tourism development till 2035 " as "... reaching the level of world leaders in the development of digital infrastructure and services, the development of digital platforms for promoting tourism products and brands, digital navigation aids and the formation of a tourist product" [41]. The strategy proposes groups of concepts associated with digitalization: digital technologies, digital solutions, and digital services. The analysis of scientific publications allowed determining the basic term-concepts that define the processes of digitalization in tourism: 'digital tourism', 'e-tourism', and 'smart tourism' [22, 25, 26, 28, 31, 42].

eTourism is just one of the outcomes of the tourism industry incorporation of technology [42]. Smart tourism is connected with devices generating big data of various nature for monitoring tourist behavior, tourism management, and tourism marketing [25]. Digital tourism is the convergence between the physical and digital worlds, supported by sensors that collect data resulting from the interaction of tourists and the environment [31].

Studies in the area of digital, electronic, and smart tourism stay relevant. Most foreign analysts consider the results to be insufficient, reflecting only the most noticeable aspects of the ongoing digital transformations.

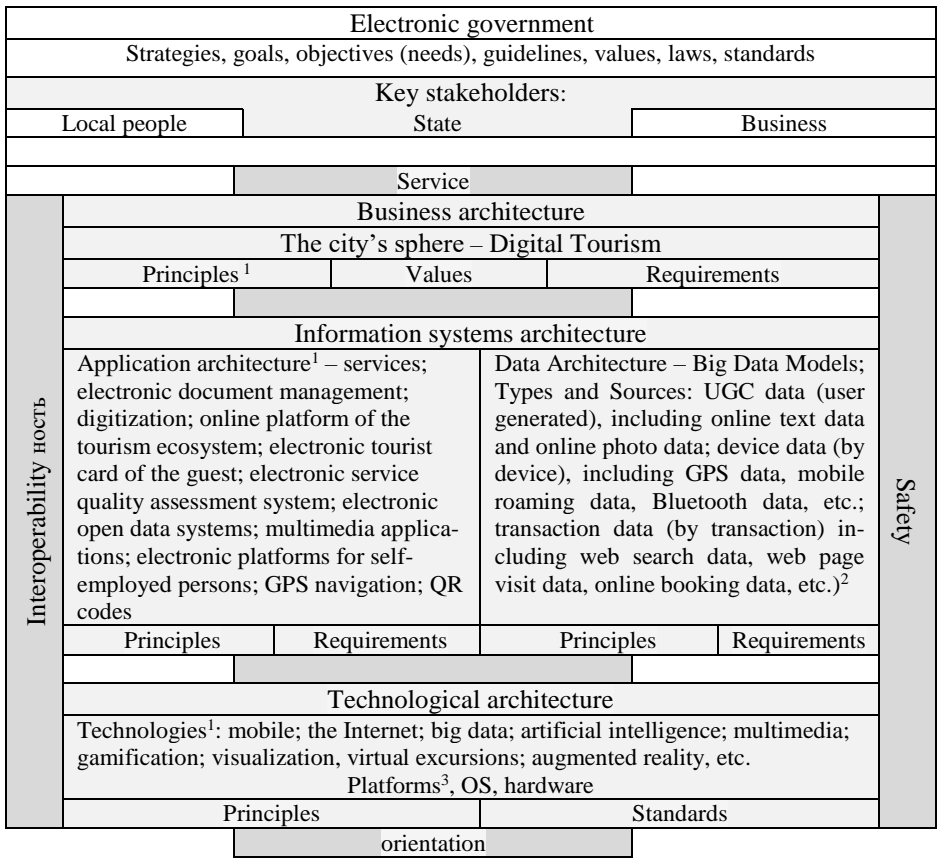

**Fig. 1.** Metamodel of smart city architecture: tourism [3, 26, 41].

The solution to urban problems can be found through the use of smart city concepts and projects. This decision supports public administration oriented towards the interests of citizens [8, 30]. The implementation of smart city (region) projects requires an architectural approach [1, 7, 43] and involves obtaining a number of results: concepts, principles, architecture (models) of a smart city, as well as areas of urban economy. The technology is a key moment in the smart city development [18]. Figure 1 shows a simplified metamodel of smart city architecture based on the Open Group TOGAF 9.2 standard [34] as applied to digital tourism as one of the urban economy areas. The TOGAF standard takes a holistic approach to enterprise architecture that allows us to see the individual blocks and elements in logic and relationships. The proposed architecture reflects an integrated view of digital tourism that allows classifying studies in this area as interdisciplinary research. Moreover, architecture is a way to identify semantic groups, which will be the base to build the trends reflected the topic and the development dynamic of interdisciplinary direction. M. Kohno, a researcher from Japan, [16] has studied concepts and implementations of smart cities, basing on the main trends characterizing smart city models around the world. As a result of this monitoring, Kohno has concluded that a modern smart city is not just focused on certain technology packages. In fact, this is the city which is a way to create and shape a new society, to solve a problem that limits the life of this community, and to determine a breakthrough into the future.

At the moment, BigData, the Internet of Things (IoT), augmented reality, artificial intelligence, blockchain technologies, as well as mobile and Internet technologies have become widespread in the tourism sector of the smart city. Artificial intelligence is used to automate aircraft servicing, improve profitability management for hotels and airlines, personalize booking mechanisms, manage airport traffic, and operate chat bots on travel company websites. Technologies of virtual and augmented reality (AR/VR) in the tourism industry have found their application in navigation, information support for the management and organization of excursion activities, and virtual travel. The relevance of sources and data are provided by the technologies of the Internet of Things. For instance, sensors in a smart hotel can collect a huge amount of data, which use can improve the quality of travel products and personalize travel services. The use of mobile devices has become a major trend in the travel business in recent years. That is why the analysis of contexts and the construction of trends in digital tourism studies is becoming important.

## 2. Methods used in interdisciplinary research areas

As noted earlier [19], in modern scientific studies, the task of analyzing promising interdisciplinary scientific areas is being actualized, which allows us to predict the relevance of research results in these areas (knowledge) in various fields of social life. There is a lag in the development of scientometric and scientific disciplines from the growth rate of the terminological and categorical base of interdisciplinary scientific fields, which is uncontrollably formed by scientific schools, groups, and individual researchers. The ambiguity of terminology and unstructured information, even with free access, makes it impossible to quickly monitor emerging trends and relationships.

Reinterpretation the role of interdisciplinary disciplines and the applicability of traditional scientific methods in interdisciplinary research is widely covered in scientific publications. Interdisciplinary areas (in fact contexts) are believed to be identified through a thematic search and described quantitatively and qualitatively through various kinds of measurements and procedures. Thus, for instance, K. Okamura, focusing on clusters of highly cited papers widely known as research fronts (RFs), has suggested that interdisciplinarity is statistically significantly and positively associated with research impact by focusing on highly cited paper clusters [32]. C. Carusi and G. Bianchi have applied the quantifying of journals' interdisciplinarity by exploiting relation between scholars and journals where such scholars do publish [4]. J. Raimbault has offered a measurement methodology that combines the analysis of citation networks and semantic analysis, studying the qualitative laws of relations between endogenous disciplines [40]. C. Piciocchi and L. Martinelli have conducted research on the development of categorical-conceptual apparatuses of various scientific areas, where Digital Humanities methods and approaches are used [38]. The paper on the cultural initiatives and the mapping process of governance and participation dynamics involve an abductive reasoning as a pragmatic approach to advancing the social sciences through a process of "systematic combining" both the analytical framework and the case studies in order to expand the theoretical and empirical understanding of results [2]. The methodology of research groups [14, 35] consists of data extraction, analytical approach (descriptive analysis and text mining), and machine-learning analysis. G. Paré, M. C. Trudel, M. Jaana, S. Kitsiou focus the researchers on the preparation as well as analysis of reviews, and develop their own typology of reviews, encouraging authors to use this typology to position their contribution [34]. Gamification studies constitute a significant part of such studies. For instance, Finnish researchers practice intellectual search and analysis of scientific texts in their study, preferring manual data processing [17]. The search strategy of Indonesian authors described in their paper "The Approach to the Meta-description of the Interdisciplinary Research Terminological Landscape" consists of "sorting digital libraries, determining keywords, using existing tools in digital libraries to facilitate the search, and taking primary studies obtained for processing" [39].

The review of research approaches presented below was performed on the selection of articles in the Science Direct library as a result of the complex request that includes the main term-concepts of the area: 'e-tourism', 'smart tourism', and 'digital tourism'.

A generalization of the analytical methods used in the research of digital and smart tourism, which belongs to interdisciplinary directions is presented in the paper "Big data in tourism research: a literature review" by Jingjing Li and et al. [25]. The paper notes that a variety of text mining techniques, which include data collection and analysis, are widely used in tourism research to extract and apply useful information hidden in online text data. The online text data collected by a web scanner is analyzed to extract useful knowledge (context knowledge) in two stages: data preprocessing and pattern detection.

Preprocessing includes data cleansing, tokenization, word wrap, and part-of-speech marking operations. Web search results are used to predict tourism trends. For this, two main steps are taken: the choice of keywords and the introduction of predictors, which allows you to build a predictive model in the future. The selection of keywords (the terminological core of the scientific direction) is the main process in tourism research by using web-search data, and the results highly depend on the selection methods.

The study by Julio Navío-Marco et al. [31], according to the authors, can be classified as a narrative review aimed at analyzing the academic literature linking ICTs and tourism as well as making a critical assessment of its quality. The review was conducted in a traditional, namely, conceptual and chronological manner. From a methodological point of view, the online databases of scientific publications (Web of Science, ScienceDirect) and various combinations of keywords related to e-tourism were used. Since in a number of previous publications the predominant use of scientific articles on tourism was declared as the main limitation of research (which would not correspond to the interdisciplinary nature of the topic under consideration), the authors included journals related to ICTs in the review. In the study, the following review procedure was adopted: at the beginning, the determination of the purpose and scope (period, subject of the review); then the determination of the selection materials' procedure; the adjusting of the source selecting process (manual search and identification of relevance using various combinations of key-words, inclusion of JCR Q1 articles with a significant number of citations in the review); the compilation of the most relevant materials for analysis; the reselection and the review of sources; highlighting of the relevant content with using the multiple coders. Having given the presence of a large number of topics and approaches reflecting the research of the links between ICTs and tourism in recent years, the authors of the paper argue that it was impossible to imply an overview of such a multi-dimensional area as 'smart tourism' without limiting any volume of material.

Jing Li et al. [24] have applied such a research methodology as the complex analysis of critical media discourse (CMDA), which is the integration of the analysis of media discourse Carvalho (MDA, 2008) with Fairclough Critical Discourse Analysis (CDA, 1995) methodology research. The CDA approach provides a skeleton framework; while the media discourse analysis framework suggests analytical components for each CDA dimension. CDA includes three various types of analysis: text analysis, processing analysis, and social analysis, which are simultaneous but interdependent. CDA is suitable for the study of social and cultural change because it facilitates the integration of discourse analysis and analysis of macro-contexts. The Carvalho text analysis scheme was applied to the text description. Contextual analysis was used to interpret the processes. The dataset in this work is news, industry reports and reviews, magazine articles, expert reviews, travel advertisements, editorials, and travel notes collected through the Google search engine. The research was conducted using Leximancer software to identify the main concepts and dominant themes in media discourses, followed by manual coding for CMDA. Data collection began with the selection of keywords. As the basic term-concepts and keywords for collecting data from media discourses, the combinations that most often appeared in scientific research were chosen. The definition of basic term-concepts and their relationships is an integral part of discourse analysis, and the vocabulary used to represent a certain phenomenon is an essential component for the explication of meanings. Close attention was paid to the formation of the thesaurus, including the term-concepts found in the headings and first paragraphs of articles in the media. As a result, 94 term-concepts were identified in the work, which were divided into twelve semantic topics. The thesaurus reflected the semantic diversity of the contexts of the materials processed.

The aim of the systematic literature review presented by Sanaz Shaee et al. [42] was to explore various aspects of intelligent travel destinations. A systematic review is an explicit and comprehensive method for identifying, synthesizing and evaluating, and consolidating the results of existing research on a specific topic of interest to a wide range of researchers. The review used a grounded theory method aimed at exploring a specific phenomenon through an inductive process that generates a theoretical understanding of the phenomenon. The approach is useful for conducting a comprehensive theoretical analysis related to a topic and consists of explaining the target phenomenon in accordance with concepts, categories, and relationships between them. Data analysis begins with open coding (defining categories, sentences, and dimensions), continues with axial coding (examining strategies, conditions, and consequences), and ends with selective coding (theory generation).

Primary links were selected based on an overview of headings, keywords, and abstracts. The queries were generated using the following keywords: smart tourism, smart travel destinations, smart city, smart sustainable city, information and communication technology, and sustainable tourism. Then the titles, abstracts, and keywords of the articles were revised in order to find other terms and keywords used in the research literature and to develop their own set of keywords. Boolean operators were used to get the best search results: logical AND to combine basic terms and OR to include synonyms.

Thus, the research environment has developed certain approaches and methods aimed at studying contextual knowledge. Without abandoning the generic and most adequate procedures, in our study we propose to use a synthetic method, which involves the use of ICTs at all stages of the study: from the search and collection of information to its quantitative and qualitative processing. Scientific text analysis is based on technologies of contextual search, explication, and display of contextual knowledge (contexts). The paper demonstrates the results of applying this method.

## 3. Research Methodology

The methodology of that research is a comprehensive approach developed by the authors (named as synthetic method). There are several attitudes inherent in Synthetic method. The attitudes are the principles of conducting research, which include:

— synthesis of various search methods, integrated coverage of research tools and varying the sequence of application of search technologies, selection, explication and analysis of contextual knowledge, depending on the initial conditions and characteristics of a specific research;
— selection of digital resources containing text arrays, reflecting both scientific and socio-political discourses;
— refusal to study a thematic sample of highly cited scientific journals with a high impact factor in favor of considering a wider range of publications from thematically different editions, which makes it possible to explicate a larger number of rel-evant terms, as well as to consider various trends in the development of interdisciplinary research areas, regardless of their prevalence.

The application of the synthetic method is independent of the choice of specific information environments and software.

## 4. Dynamics of research interest in "Digitalization of tourism" interdisciplinary direction

The results obtained at the previous stage of identifying term-concepts in the subject area of digital tourism were used to analyze the dynamics of research interest in this topic both in the world and in Russian scientific discourses.

A contextual search was made in scientific network information resources and mass media to form a Russian-language terminological base. Search queries have included the key phrases such as 'digital tourism', 'digital tourism', 'smart tourism', 'smart tourism', 'e-tourism', 'e-tourism', 'intellectual tourism' (in Russian and English). The inclusion of English terms in search queries is explained by the fact that the authors of articles often use both Russian and English versions of terms. Figure 2 presents the dynamics of publication activity on the topic of digital and e-tourism, demonstrated on the basis of the Russian scientific electronic library eLibrary and the Google Scholar information retrieval system from 2010 to 2020.

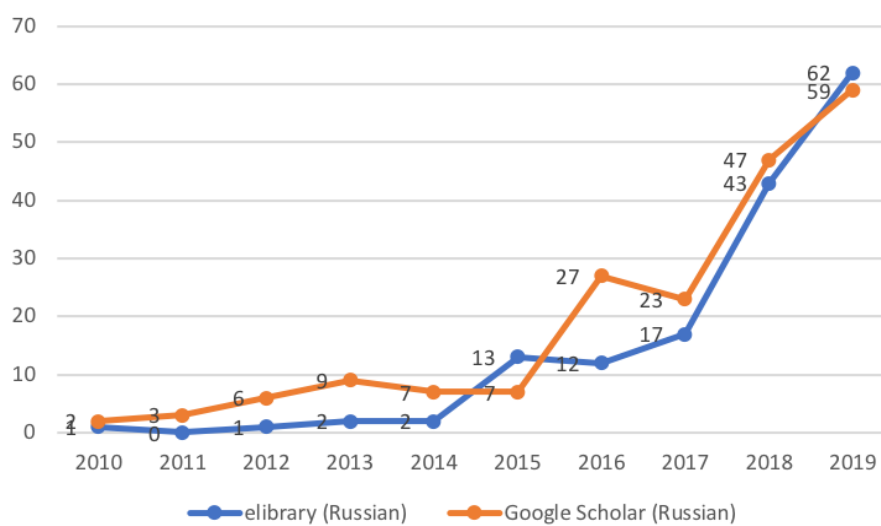

**Fig. 2.** Dynamics of research interest in digital tourism 2010-2020 in Russian-language scientific discourse

Analysis of the dynamics of the Russian-language terminology base of digital tourism, obtained from the eLibrary and Google Scholar, allowed us to draw the following conclusions:

— the activity of scientific interest in the topic of digital tourism has been growing over a ten-year period. For example, in 2019, the activity increased relative to the study on eLibrary by 60 times, on Google Scholar by 30 times;

— trends expressed in scientific publications and formed on the basis of the eLibrary platform and Google Scholar show the same character;

— the general trend can be divided into two blocks. the first block of "quiet" is the period from 2010 to 2015 (the number of publications is insignificant and varies within 1-7, there are no abrupt changes) and the period of "increased activity" is the period after 2015 to the present (the number of publications is growing rapidly from year to year and varies in the range of 13-62). Moreover, compared to 2015, the activity on eLibrary increased by 4 times, on Google Scholar by more than 8 times.

Similar data was obtained from Google Scholar, ScienceDirect, and SpringerLink. The choice of ScienceDirect and SpringerLink platforms is due to the fact that they represent scientific publications of the largest scientific publishers Elsevier and Springer. the query was used with the use of the main terms and concepts identified at the previous stage: eTourism, smart tourism, digital tourism, intelligent tourism. the data obtained are shown in figure 3.

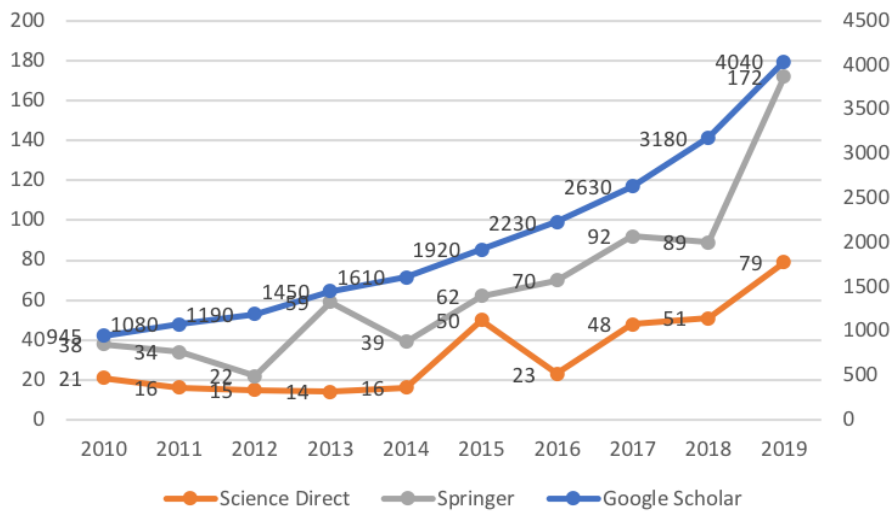

**Fig. 3.** Dynamics of research interest in digital tourism 2010-2020 in the English-language scientific discourse

From the Integrum information system results were obtained from 2010 to 2019 for Russian-language queries: ("digital tourism" or "electronic tourism") and ("smart tourism "or" smart tourism"). The results obtained indicate a rather weak interest of the official media in the development of the tourism sector based on the use of information and communication technologies. So, in the period from 2010 to 2019, only 18 publications were identified for the first request, and 17 for the second, with a clearly chaotic spread over the years. The discourse in mass media shows that journal-ists use the terminology of "digital tourism" poorly. At the same time, they do not bother to seriously immerse themselves in this topic. Basically, the term "digital tourism" or "smart tourism" is used once in the title or annotation, and then no terms reflecting the topic of "digital tourism" are found in the text.

In addition, the impact of the COVID-19 coronavirus pandemic, which began at the end of 2019, on ongoing research in the development of "digital tourism" was analyzed. To do this, the terms COVID, SARS-CoV, 2019-nCoV and coronavirus were added to the queries in Russian and English in Google Scholar:

```
eTourism OR "smart tourism" OR "digital tourism" OR "in-
telligent tourism" +(COVID OR SARS-CoV OR 2019-nCoV OR
coronavirus)
```

Publications with these terms were found only in 2020, which is justified by the impact of the pandemic on the tourism industry on a global scale not earlier than march 2020. Analysis of the obtained data shows (fig. 4) that:

— in the total flow of publications on digital tourism, about 15 percent cover the development of digital tourism in the new conditions associated with the spread of COVID-19;
— Russian-speaking researchers are in the trend of global trends in the development of digital tourism;
— the Russian tourism industry has suffered to the same extent as the global tourism industry, and digital tourism is the driver of its development in the context of the COVID-19 coronavirus epidemic.

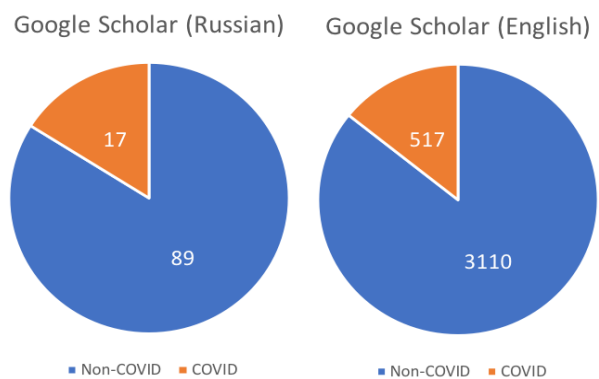

**Fig. 4.** The correlation between publications on digital tourism in general and those dealing with the problems of COVID-19 (Russian and world scientific discourse 2020)

It has not been long since the beginning of the COVID-19 coronavirus pandemic. therefore, the dynamics of its impact on research in the field of digital tourism can be traced in a few years, if the current situation with the coronavirus continues to exist or will have trends for further development.

## 5. The digital solutions of St. Petersburg

These days, digital technologies are deeply embedded in the tourism industry. However, a few years ago, their functions in tourism were reduced mainly to information and communication functions. Tourism in the world has a single information space based on new principles of information support and tourism management. Customer-centricity is the hallmark of smart travel destinations. The main task of this centricity is to provide services which can attract more tourists and increase comfort. For that, special tourist applications were created. Table represents the implementation of digital travel services for the two cities (Tel Aviv and St. Petersburg).

**Table 1.** Implementation of digital travel services of Tel Aviv and St. Petersburg.

| Smart travel destinations | Functions | Components | Travel applications | |
|---|---|---|---|---|
| | | | Tel Aviv | St. Petersburg |
| Augmented reality | allow tourists to combine real experiences and virtual information | Smart people | "Mekomi" – video guide to attractions | audio guide to the city & the Hermitage |
| Vehicle tracking system | provide information on schedules, tariffs, routes, transport stops, ordering a taxi in real time | Smart mobility | "Alternative" – information on how to move | Yandex.transport – maps, routes |
| Online booking and purchase | rental housing, transport, excursions comfort and cost savings, familiarity with local customs and culture | Smart economy | "Tellavista" – comprehensive booking system | "Gettable" – online restaurant reservations |
| Joint consumption | exchange vacation | Smart economy, Smart living | "Casa Versa" – exchange vacation platform | "House Exchange" – exchange vacation platform; Yandex.Drive, Delimobil, Velogorod, Whoosh and others |
| Loyalty programs | services with discounts | Smart economy, Smart mobility | "MyeFlyTM" – purchase of airline tickets for loyalty miles | "Guest card" – a single card for tourists, which includes visits to museums and excursions, electronic transport ticket, discount coupons |
| Access system to officials | registration of complaints, appeals, dialogue by the authorities and regulatory organizations | Smart government | "Tel Aviv Municipality" – communication with the city administration | "State services portal St. Petersburg" – learning state and municipal services, electronic online applications |

Analysis of the presented data shows that the tourism industry uses digital solutions in different countries. At the same time, large cultural centers are in the trend of global development. this is also typical for Russia.

The use of mobile devices has become a major trend in the travel business in recent years. On average, tourists use 17 different mobile apps while traveling [33]. First of all, these are cartographic services, travel guides and audio guides, geolocation and geoinformation systems, booking air and railway tickets, hotels, various types of reference information, translation programs, and others.

St. Petersburg has gained mobile applications and services, which help tourists to navigate and make their staying in the city more comfortable and variable. Visit Petersburg.ru is a city tourist portal, which has a large database of objects of the city tourist industry in nine languages. There are other guide apps for St. Petersburg: *Peterburg.center, izi.TRAVEL, KudaGo, TripAdvisor*, and others. Application-guides to museums and sights are *Hermitage, Kunstkamera. museum guide*; audio guides – *Erarta 2.0, RM Guide* (Russian museum), 1000Guides (St. Petersburg museums). AR/VR technologies in St. Petersburg are represented by the following series: mobile application *Museum of AR Streets,* VR film *Hermitage VR. Immersion in History*, a virtual tour of the Russian Railways Museum, VR bus tour *I see the city of Petrov*. AR/VR technologies allow providing cultural leisure for all categories of citizens even in a pandemic. Since the end of 2019, electronic tourist visas have begun to operate in St. Petersburg and the Leningrad region.

## 6. Conclusion

Implemented study allows us to draw the following conclusions:

— digital tourism is perceived in society as a new and still poorly studied phenomenon without a clear understanding of the essence of this concept;
— in the educational sphere, new areas of training of specialists and advanced training of civil servants in the field of digital tourism appear;
— initiative projects appear and begin to be implemented (mainly at the regional level) in the mainstream of digital tourism and smart tourism (to a greater extent), for example, various specialized Internet portals, aggregators, and search engines;
— digital tourism and smart tourism are considered as mechanisms for attracting people to the sphere of real tourism (implementation of participation mechanisms);
— discussion of the digital and smart tourism development occurs at various regional and international venues (forums and festivals);
— In this way, the data illustrating publication activity allows asserting a growing scientific interest in the digital tourism topic.

The development of digital technologies, the ubiquity of the Internet, the widespread development of social networks, the development of technologies related to artificial intelligence, machine learning and BigData, and AR / VR have a great impact on the tourism industry transformation. The use of these technologies increases the efficiency of communication between participants in the tourism market. Digital transformations in the tourism sector of St. Petersburg have a large scope: mobile applications, the use of augmented reality technologies, audio guides, a single portal for tourists, and many others.

Future research initiatives could include questions about how technology fits into public conscience and relationships, and how tourism-related businesses are transformed. An equally important area is scientific research, tracking scientific trends, and building their predictive models. Such research is of value to both scientists and businessmen. This is especially true in the light of forecasting the development of tourism after the end of the pandemic.

## Acknowledgement

The work was supported by the Russian Foundation for Basic Research, project 18-011-00923-a.

## References

1. Ahlund, C.: ICT Architectures for Smart Cities/Regions. https://www.kth.se/ polopo-ly_fs/1.582402!/C%C3%85_ICTArchitectures_for_SmartRegions.pdf, last accessed 2020/12/10.
2. Biondia, L., Demartinia, P., Marchegiania, L., Marchioria, M., Piberb, M.: Understanding orchestrated participatory cultural initiatives: Mapping the dynamics of governance and participation. Cities 96, 102459 (2020). DOI: 10.1016/j.cities.2019.102459.
3. Buhalis, D., Licata, M.C.: The future eTourism intermediaries. Tourism Management 23, 207–220 (2020).
4. Carusi, C., Bianchi, G.: A look at interdisciplinarity using bipartite scholar/journal networks. Scientometrics 122, 867–894 (2020). DOI: 10.1007/s11192-019-03309-3.
5. Cherevichko, T.V., Temjakova, T.V.: Cifrovizacija turizma: formy projavlenija. Izv. Sarat. un-ta. Nov. ser. Ser. Jekonomika. Upravlenie. Pravo, 19(1), 59–64 (2019).
6. Chetvertaja promyshlennaja revoljucija. Populjarno o glavnom tehnologicheskom trende XXI veka, TAdviser 17.10.2017 (2017),
http://www.tadviser.ru/index.php/Статья:Четвертая_промышленная_революция_(Indus try_Индустрия_4.0), last accessed 2020/11/11 [In Russian].
7. Creating Municipal ICT Architectures. A reference guide from Smart Cities. https://www.slideshare.net/smartcities/creating-municipal-ict-architectures-a-reference-guide-from-smart-cities (2011), last accessed 2020/12/10.
8. Drozhzhinov, V.I., Kupriyanovsky, V.P., Namiot, D.R., Sinyagov, S.A., Kharitonov, A.A.: Smart Cities: models, tools, rankings, and standards. International Journal of Open Information Technologies 5(3), 19–48 (2017).
http://injoit.org/index.php/j1/article/view/403, last accessed 2020/12/10.
9. Egger, I., Lei, S.I., Wassler, P.: Digital free tourism – An exploratory study of tourist motivations. Tourism Management 79, 104098 (2020).
DOI: 10.1016/j.tourman.2020.104098.
10. EU Guidebook on Sustainable Tourism for Development. Enhancing capacities for Sustainable Tourism for development in developing countries, Madrid, UNWTO (2013). https://www.unwto.org/EU-guidebook-on-sustainable-tourism-for-development., last accessed 2020/12/10.
11. Federal'nyj zakon ot 24.11.1996 N 132-FZ (red. ot 03.07.2019, ot 01.04.2020). "Ob osnovah turistskoj dejatel'nosti v Rossijskoj Federacii",
http://www.consultant.ru/document/cons_doc_LAW_12462/bb9e97fad9d14ac66df4b6e67 c453d1be3b77b4c/, last accessed 2020/12/10 [In Russian].
12. Gradinarova, A.A.: Sovremennye tendentsii tsifrovoy transformatsii v turisticheskoy otras-li. Trudy konferentsii «Problemy i perspektivy razvitiya turizma v Yuzhnom federal'nom okruge». Izd-vo «Tipografiya «Arial», Simferopol', pp. 69–73 (2017).

13. Holt, J., Polton, J., Huthnance, J., Wakelin, S., Enda, O'Dea E., Harle, J., Yool A., Artioli Y., Blackford Y., Siddorn J., Inall, M.: Climate-Driven Change in the North Atlantic and Arctic Oceans Can Greatly Reduce the Circulation of the North Sea. Geophysical research Letters (2018), DOI: 10.1029/2018GL078878.

14. Jimenez-Marquez, J.L., Gonzalez-Carrasco, I., Lopez-Cuadrado, J.L., Ruiz-Mezcua, B.: Towards a big data framework for analyzing social media contente. International Journal of Information Management, 44, 1–12 (2019). DOI: 10.1016/j.ijinfomgt.2018.09.003.

15. Kalmakova, A.A.: Cifrovye turisticheskie jekosistemy i ih rol' v marketinge destinacij. Geografija i turizm: sb. nauch. tr. 14, 57–62 (2015) [In Russian].

16. Kohno, M.: Innopolis is an outdated model, it should have been implemented 30 years ago (2018). https://realnoevremya.ru/articles/95516-intervyu-s-michinaga-kohno-ekspertom-po-umnym-gorodam, last accessed 2020/12/10.

17. Koivisto, J, Hamari, J.: The rise of motivational information systems: A review of gamification research. International Journal of Information Management, 45, 191–210 (2019). DOI: 10.1016/j.ijinfomgt.2018.10.013.

18. Kononova, O.V., Pavlovskaya, M.A.: Digital economy technologies in smart city projects: participants and prospects. Modern Information Technologies and IT-Education, 14(3), 692–706 (2018). DOI: 10.25559/sitito.14.201803.692-706.

19. Kononova, O., Prokudin, D., Timofeeva, A.: Gamification and interdisciplinary scientific research: scientific text mining. In: Proceedings of the international conferences on ICT, Society and Human Beings 2020, Connected Smart Cities 2020 and Web Based Communities and Social Media 2020, july 21 - 23, 2020. Edited by Piet Kommers and Guo Chao Peng, IADIS Press, pp. 209-213 (2020).

20. Kononova, O.V., Lyapin, S.Kh., Prokudin, D.E.: Studying the Interdisciplinary Terminological Landscape of Digital Economy with the Use of Contextual Analysis Tools. International Journal of Open Information Technologies, 6(12), 57–66 (2018), http://injoit.org/index.php/j1/article/view/648, last accessed 2020/11/11 [In Russian].

21. Kononova, O.V., Prokudin, D.E., Smirnova, P.V.: Approach to Use of Network Scientific Environment for Studying the Interdisciplinary Terminological Landscape of Digital Economy. Information Society: Education, Science, Culture and Technology of Future, 3, 53–66 (2019). DOI: 10.17586/2587-8557-2019-3-53-66 [In Russian].

22. Kontogianni, A., Alepis, E.: Smart tourism: State of the art and literature review for the last six years. Array 6, 100020 (2020). DOI: 10.1016/j.array.2020.100020.

23. Kormjagina, N.N.: Smart-turizm kak chast' Smart-koncepcii. Marketing i logistika, 6(14), 45–57 (2017) [In Russian].

24. Li J., Pearce P.L., Low D.: Media representation of digital-free tourism: A critical discourse analysis. Tourism Management, 69, 317–329 (2018). DOI: 10.1016/j.tourman.2018.06.027.

25. Li, J., Xu, L., Tang, L., Wang, S., Li, L. Big data in tourism research: A literature review. Tourism Management, 68, 301–323 (2018). DOI: 10.1016/j.tourman.2018.03.009.

26. Li, Yu., Hu, C., Huang, Ch., Duan, L.: The concept of smart tourism in the context of tourism information services. Tourism Management, 58, 293–300 (2017). DOI: 10.1016/j.tourman.2016.03.014.

27. Molchanova, V.A.: Tendencii innovacionnogo razvitija turistskih destinacij: «umnaja destinacija». Jekonomika i predprinimatel'stvo, 9(3), 715–720 (2017) [In Russian].

28. Molz, J.G.: Travel connections: Tourism, technology and togetherness in a mobile world. London: New York, Routledge (2012).

29. Moshnjaga, E.V.: Osnovnye tendencii razvitija turizma v sovremennom mire. Vestnik RMAT, 3(9), 20–33 (2013) [In Russian].

30. Narmeen, Z.B., Jawwad, A.S.: Smart City Architecture: Vision and Challenges. International Journal of Advanced Computer Science and Applications, 6(11), 246–255 (2015). http://thesai.org/Downloads/Volume6No11/Paper_32-Smart_City_Architecture_Vision_and_Challenges.pdf, last accessed 2020/12/10.

31. Navío-Marco, J., Ruiz-Gómez, L.M., Sevilla-Sevilla, C.: Progress in information technology and tourism management: 30 years on T and 20 years after the internet – Revisiting Buhalis & Law's landmark study about eTourism. Tourism Management, 69, 460–470 (2018). DOI: 10.1016/j.tourman.2018.06.002.

32. Okamura, K.: Interdisciplinarity revisited: evidence for research impact and dynamism. Palgrave Communication, 5, 141 (2019). DOI: 10.1057/s41599-019-0352-4.

33. Opros tsifrovykh puteshestvennikov Rossii. https://t.co/d8jRGAxnnh, last accessed 2020/12/10.

34. Paré, G., Trudel, M.C., Jaana, M., Kitsiou, S.: Synthesizing information systems knowledge: A typology of literature reviews. Information & Management, 52(2), 183–199 (2015). DOI: 10.1016/j.im.2014.08.008.

35. Pejic-Bacha, M., Bertoncelb, T., Meškob, M., Krstićc, Ž.: Text mining of industry 4 job advertisements. International Journal of Information Management, 50, 416–431 (2020). DOI: 10.1016/j.ijinfomgt.2019.07.014.

36. Perles Ribes, J.F., Baidal, J.I.: Smart sustainability: a new perspective in the sustainable tourism debate. Investigaciones Regionales – Journal of Regional Research, 42, 151–170 (2018).

37. Phillips, S.G.: The tourism industry association of Canada [EB/OL], http://www.slideshare.com, last accessed 2020/12/10.

38. Piciocchi, C, Martinelli, L.: The change of definitions in a multidisciplinary landscape: the case of human embryo and preembryo identification. Croatian Medical Journal, 57(5), 510–515 (2016). DOI:10.3325/cmj.2016.57.510.

39. Purwandari, B., Sutoyo, M.A.H., Mishbah, M., Dzulfikar, M.F.: Gamification in e-Government: A Systematic Literature Review. In: Fourth International Conference on Informatics and Computing (ICIC), pp. 1–5. Semarang, Indonesia (2019). DOI: 10.1109/ICIC47613.2019.8985769.

40. Raimbault, J.: Exploration of an interdisciplinary scientific landscape. Scientometrics, 119(2), 617–641 (2019). DOI: 10.1007/s11192-019-03090-3.

41. Rasporjazhenie Pravitel'stva RF ot 20.09.2019 № 2129-r. "Ob utverzhdenii Strategii razvitija turizma v Rossijskoj Federacii na period do 2035 goda". http://static.government.ru/media/files/FjJ74rYOaVA4yzPAshEulYxmWSpB4lrM.pdf, last accessed 2020/12/10 [In Russian].

42. Shaee, S., Ghatari, A.R., Hasanzadeh, A., Jahanyan, S.: Developing a model for sustainable smart tourism destinations: A systematic review. Tourism Management Perspectives, 31, 287–300 (2019). DOI: 10.1016/j.tmp.2019.06.002.

43. Shelton, T., Zook, M., Wiig A.: The actually existing Smart City. Cambridge Journal of Regions, Economy and Society, 8(1), 13-25 (2015). DOI: 10.1093/cjres/rsu026.

44. Shhedrina, E.Ju., Moiseeva, A.G., Goncharov, A.N., Hubulova, V.V.: Cifrovoj turizm: kak industrija 4.0 povlijaet na turisticheskuju otrasl' regiona. Sovremennaja nauka i innovacii, (1), 251–256 (2019) [In Russian].

45. Sjaocjan', K., Shucin, A.: Issledovanie razvitija «umnogo» turizma v provincii Czjansi v ramkah koncepcii «Internet+». Jekonomicheskie i social'nye peremeny v regione: fakty, tendencii, prognoz, 4, 199–205 (2016) [In Russian].

46. Smirnov, A.V., Ponomarev, A.V., Levashova, T.V., Teslja, N.N.: Podderzhka prinjatija reshenij v turizme na osnove cheloveko-mashinnogo oblaka. Iskusstvennyj intellekt i prinjatie reshenij, 2, 90–102 (2017) [In Russian]. /31

47. The Open Group. TOGAF 9.2. The Open Group, http://pubs.opengroup.org/architecture/togaf9-doc/arch, last accessed 2020/12/10.

48. Tourism the year 2000 and beyond qualitative aspects, Madrid, OMT. Madrid (1993) /21.

49. UNWTO. Tourism resilience committee stresses need for "Smart Tourism" [EB/OL]. 2012-03-11 (2012), http://www.slideshare.com, last accessed 2020/12/10.

50. Ustinova, M.V., Shevchenko, M.V.: Industrija gostepriimstva v jepohu cifrovizacii. Jepoha nauki, 20, 459–463 (2019) [In Russian].

51. Voronkova, L.P. WWW as an Information Resource for International Tourism. Information Society, (2), 49–53 (2020), http://infosoc.iis.ru/article/view/181, last accessed 2020/12/10.